\documentclass[usenatbib,useAMS,referee]{biom}
\usepackage{float}
\usepackage{multirow}
\usepackage{multicol}
 \usepackage{graphicx}
\usepackage{mathtools}
 \usepackage{listings}
 \usepackage{color}
 \usepackage{setspace}
 \usepackage{pdfpages}
 \usepackage{amsfonts}
\usepackage{amsmath}
\usepackage{amssymb}
\usepackage{subfigure}
 \usepackage{tabularx}
\usepackage{bm}
  \usepackage{url}

\newcommand{\N}{\mbox{N}}

\newcommand{\RR}{\mathbb{R}}
\newcommand{\GG}{\mathcal{G}}
\newcommand{\HH}{{\mathcal{H}}}

\renewcommand{\th}{\theta}
\newcommand{\kk}{^{(k)}}
\newcommand{\Ak}{\Ab\kk}
\newcommand{\Bk}{\Bb\kk}
\newcommand{\Sigk}{\Sigb\kk}
\newcommand{\thk}{\thb\kk}
\newcommand{\sigk}{\sigma\kk}
\newcommand{\Gk}{\GG_k}

\newcommand{\thb}{\bm{\th}}
\newcommand{\pib}{\bm{\pi}}
\newcommand{\yb}{\bm{y}}
\newcommand{\xb}{\bm{x}}
\newcommand{\ab}{\bm{a}}
\newcommand{\bb}{\bm{b}}
\newcommand{\Ab}{\bm{A}}

\newcommand{\Ib}{\bm{I}}
\newcommand{\ImA}{(\Ib-\Ab)}
\newcommand{\Bb}{\bm{B}}

\newcommand{\Sigb}{\bm{\Sigma}}
\newcommand{\Omegab}{\bm{\Omega}}

\newcommand{\at}{\tilde{a}}

\newcommand{\bt}{\tilde{b}}
\newcommand{\atk}{\tilde{a}\kk}
\newcommand{\btk}{\tilde{b}\kk}
\newcommand{\ak}{a\kk}
\newcommand{\bk}{b\kk}

\usepackage{xcolor}         

\newcommand{\suppl}[1]{Web Appendix #1}

\title[hRGM]{Heterogeneous Reciprocal Graphical Models}

\author{Yang Ni$^{1,*}$\email{yangni87@gmail.com}, 
Peter M\"uller$^{2}$, Yitan Zhu$^{3}$, and Yuan Ji$^{3,4}$ \\
$^{1}$Department of Statistics and Data Sciences, The University of Texas at Austin \\
$^{2}$Department of Mathematics, The University of Texas at Austin\\
$^{3}$Program for Computational Genomics \& Medicine, NorthShore University HealthSystem\\
$^{4}$Department of Public Health Sciences, The University of Chicago}

\begin{document}
	
\label{firstpage}
\begin{abstract} 
We develop novel hierarchical reciprocal graphical models to
infer gene networks from heterogeneous data. In the case of data that
can be naturally divided into known groups, we propose to connect
graphs by introducing a hierarchical prior across group-specific
graphs, including a correlation on edge strengths across graphs.
Thresholding priors are applied to induce sparsity of the estimated
networks. In the case of unknown groups, we cluster subjects into
subpopulations and jointly estimate cluster-specific gene networks,
again using similar hierarchical priors across clusters.  
We illustrate the proposed approach by simulation studies and three
applications with multiplatform genomic data for multiple cancers.
\end{abstract}

\begin{keywords}
Dirichlet-multinomial allocation;  hierarchical model; model-based clustering; multiplatform genomic data; Pitman-Yor process; thresholding prior.
\end{keywords}

\maketitle

\section{Introduction}
We develop a heterogeneous reciprocal graphical model
(hRGM) to infer gene networks in heterogeneous populations. 
Traditional graphical model approaches
\citep{dempster1972covariance,lauritzen1996graphical,meinshausen2006high,whittaker2009graphical,wang2009bayesian,dobra2012bayesian,green2013sampling,Wang2013,peterson2015bayesian}
assume i.i.d. sampling. However, many applications to inference for biomedical data,
including the applications in this paper, 
include highly heterogeneous populations, and understanding and
characterizing such heterogeneity is an important inference
goal.
We therefore propose an approach that admits statistical inference for
potentially heterogeneous gene regulatory relationships across
\textit{known as well as unknown} groups/subpopulations. In particular, for
the case of known groups, we model the related graphs under a Bayesian
hierarchical model framework and allow the information to be shared
across different groups. Borrowing of strength is implemented for the
graph structure as well as for the edge strengths. For the
case of unknown groups, we propose to cluster the subjects into
subpopulations with meaningfully different graphs, that is, with
group-specific graphs that differ in ways that allow biologically
meaningful interpretation.  Importantly, we implement clustering based on
differences in the networks, in contrast to earlier proposed
clustering methods
\citep{dahl2006model,quintana2006predictive,lau2007bayesian,rodriguez2008nested,muller2012product,lijoi2014dependent},
which are mostly based on cluster-specific mean/location.

Our work is motivated by three cancer genomic applications. In the first two
applications, we construct gene networks for different cancer
types. Recent pan-cancer studies \citep{hoadley2014multiplatform} find
both differences and commonalities across cancers. Traditional methods that assume one homogeneous population are not suitable in this case. Novel
statistical methods accounting for data heterogeneity as well as
adaptively borrowing strength across subtypes are needed. Our third
application finds motivation in breast cancer which is molecularly heterogeneous. Current classification systems based on three
biomarkers are argued to be suboptimal as a means of directing
therapeutic decisions 
for breast cancer patients \citep{di2015new}. Improving the
classification system is particularly important and yet challenging.
This calls for better ways of clustering breast cancer patients.

Graphical models are commonly used to probabilistically model gene
regulations. We build on a less commonly used class of
graphical models,
namely reciprocal graphical models 
(RGMs, \citealt{koster1996markov}). RGMs are a flexible class of models that
allow for undirected edges, directed edges and directed cycles.
These features are important for modeling biological feedback
loops. 
RGMs strictly contain Markov random fields (MRFs) and directed acyclic
graphs (DAGs) as special cases. However, RGMs are surprisingly underused
in the biostatistics and bioinformatics literature. We prefer to use 
RGMs over other graphical models, because the inclusion of directed
edges and possible cycles is critical for the three motivating
applications.

Although graphical models for homogeneous data have been studied
extensively in the literature, 
until recently only few approaches were proposed for heterogeneous data.
When the groups are known, it is
natural to ``borrow strength" across different sample groups via a Bayesian hierarchical model for group-specific graphs. We
provide a brief review of recent Bayesian methods and discuss the need
and possibility for improvement. 
\cite{mitra2016bayesian} consider MRFs
for $K=2$ groups,
identifying one group as the 
reference group and the other as the differential group. They assign a
uniform prior to the reference graph and construct a mixture prior for
the differential graph. 
Similarly,
\cite{oates2015exact} develop a multiple DAG approach for $K>2$. They
penalize the difference between graphs based on structural Hamming
distance and utilize integer linear programming to find the posterior
mode. \cite{peterson2015bayesian} couple undirected MRF graphs by assuming an MRF prior on the edges.  Computation can be
challenging when $K$ is moderate because of an analytically intractable normalization constant in the MRF
model. One common limitation of these methods is
that they only borrow strength in the graph space, leaving the strength of selected edges (e.g. partial correlation)
to be modeled/estimated
independently. One exception is \cite{yajima2015detecting}. They
propose a multiple DAG
which correlates the edge strength across
groups. However, similarly to \cite{mitra2016bayesian}, they only
consider $K=2$ groups and inference depends on the choice of
reference/differential groups. Generalization to
$K>2$ is not straightforward. Several non-Bayesian
methods \citep{guo2011joint,danaher2014joint,mohan2014node,lee2015joint,ma2016joint} have also been developed to learn MRFs across multiple groupse.

There are few approaches for unknown groups.
\cite{rodriguez2011sparse} propose a Dirichlet process
mixture of MRFs to simultaneously cluster samples into homogeneous
groups and infer undirected relationships between variables for each
group. Similarly, \cite{ickstadt2011nonparametric} propose a
multinomial-Dirichlet-Poisson mixture of DAGs to study directed
relationships within each cluster. \cite{mukherjee2015gpu} recently
develop a GPU-based stochastic search algorithm to improve the
computation in \cite{rodriguez2011sparse}.

In this article, we propose a hierarchical extension of RGMs to
heterogeneous RGMs (hRGMs) as a model for both cases, known and unknown groups. 
When groups are known, the hRGM borrows strength across groups 
for inference on the group-specific graphs as well as 
the strength of the included edges.
For unknown groups, the hRGM clusters
a heterogeneous population into homogeneous subpopulations and allows each
cluster to have its own network.

\section{Reciprocal graphs}
\label{sec:back}
A graph $\GG=(V,E)$ consists of a set of vertices $V=\{1,\dots,p\}$
usually representing a set of random variables and a set of edges $E$ connecting these
vertices. A reciprocal graph admits both directed edges
$i\rightarrow j$ and undirected edges $i-j$. Moreover, it explicitly
allows for directed cycles, which is useful for modeling feedback
mechanisms, which are a common motif in gene networks. Markov properties
(i.e. conditional independence relationships) can be read off from the
RGM through moralization. Statistically, RGMs are a strictly larger
class of probability models than MRFs and DAGs in terms of conditional
independence \citep{koster1996markov}. 
We provide a brief review of reciprocal graphs in
\suppl{A}.

\begin{figure}[h]
	\centering
		\subfigure[]{\includegraphics[width=0.2\textwidth]{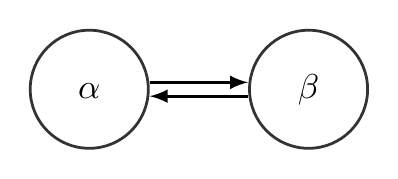}\label{tg1}} 
		\subfigure[]{\includegraphics[width=0.2\textwidth]{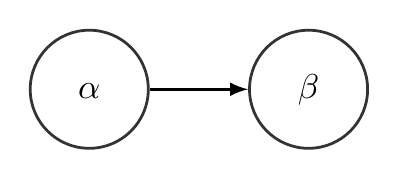}\label{tg2}}
		\subfigure[]{\includegraphics[width=0.2\textwidth]{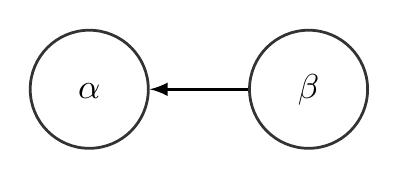}\label{tg3}} 
		\subfigure[]{\includegraphics[width=0.2\textwidth]{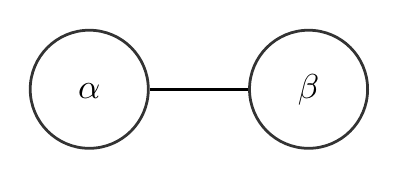}\label{tg4}}
		\subfigure[]{\includegraphics[width=0.2\textwidth]{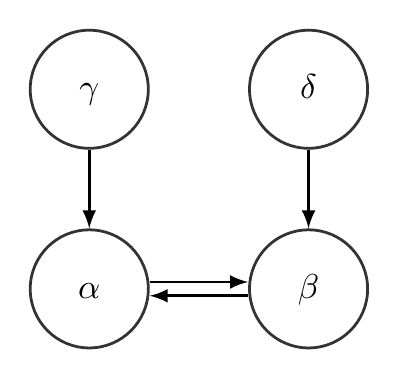}\label{rgm_fig}}
		\subfigure[]{\includegraphics[width=0.2\textwidth]{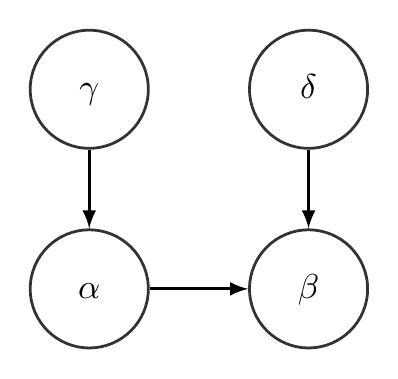}\label{dag1}}
		\subfigure[]{\includegraphics[width=0.2\textwidth]{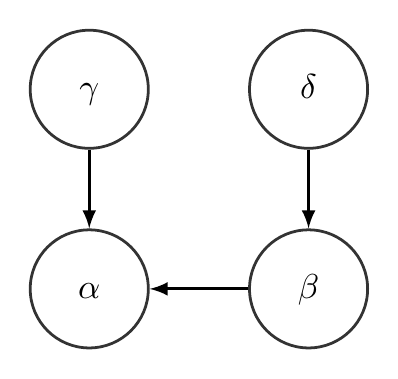}\label{dag2}} 
		\subfigure[]{\includegraphics[width=0.2\textwidth]{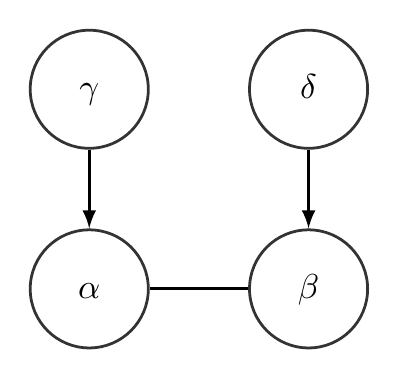}\label{ug_fig}}
	\caption{Graphs for illustration.   Graphs in (a)-(d) are
		Markov equivalent reciprocal graphs. Each graph in (e)-(h) encodes different  conditional independence relationships.}\label{rdu}
\end{figure}

RGMs are only identifiable up to a Markov equivalence class in which all
RGMs encode the same conditional independence
relationships; see a series of work by \cite{richardson1996discovery,richardson1996polynomial,richardson1997characterization} for characterization and identification of Markov equivalence class of cyclic graphs.
Beyond this
interventional or time-course data or other external information
are needed to learn the structure of RGMs and determine the directionality of the
edges. Recently, several approaches
\citep{cai2013inference,zhang2014learning,ni2016reciprocal} have 
proposed to infer directionality with observational
data by integrating multi-platform data and fixing some
edge directions based on biological knowledge.
In this paper, we follow the approach of
\cite{ni2016reciprocal} to include external information about known directional edges,
by exploiting the central dogma of molecular biology, implying that
DNA methylation and DNA copy number can affect mRNA gene expression,
but not vice versa. 
Based on this consideration, we integrate DNA copy number and
DNA methylation with mRNA gene expression and fix the direction of
potential
edges between DNA measurements and mRNA gene expressions. For
  illustration, let vertices $\alpha$ and $\beta$ in Figure \ref{rdu}
  denote mRNA gene expressions. With observational data, we can not
  differentiate between graphs \ref{tg1}, \ref{tg2}, \ref{tg3} and
  \ref{tg4} because they all define the same conditional independence
  relationships between vertices $\alpha$ and $\beta$ (i.e. none). The
  identifiability issue can be resolved by incorporating additional
  information.  Let vertices $\gamma$ and $\delta$ represent copy
  number or methylation of the corresponding genes. We introduce
  fixed directed edges $\gamma\rightarrow\alpha$ and
  $\delta\rightarrow \beta$. Then the relationship between
  vertices $\alpha$ and $\beta$ are identifiable because each of
  graphs \ref{rgm_fig}, \ref{dag1}, \ref{dag2} and \ref{ug_fig}
  encode different conditional independence relationships.
A theoretical justification for this framework is discussed in
\cite{logsdon2010gene} and 
\cite{oates2016estimating}. 

We briefly review the mapping between RGMs and 
simultaneous equation models (SEMs). Let $\yb_i\in \RR^p$
denote gene expressions for $p$ genes and $\xb_i\in \RR^q$
denote $q$ DNA level measurements (copy number and methylation in our
case study) for subject $i=1,\dots,n$. Consider $(\yb_i,\xb_i)$
to be jointly normal and satisfy the following SEM 
\begin{eqnarray}
  \label{se}
  \yb_i=\Ab\yb_i+\Bb\xb_i+\bm{e}_i
\end{eqnarray}
where $\Ab=(a_{jj'})\in \mathbb{R}^{p\times p}$ with zeros on the
diagonal, $\Bb=(b_{jj'})\in \mathbb{R}^{p\times q}$, $\bm{e}_i\sim
N_p(0,\bm{\Sigma})$ with diagonal covariance matrix
$\bm{\Sigma}=\mbox{diag}(\sigma_1,\dots,\sigma_p)$ and $\bm{e}_i$. The linearity and normality assumptions of the SEM in (\ref{se}) can be relaxed by using, for example, the approaches in \cite{mooij2011causal} and \cite{lacerda2012discovering}. Assuming that $\Ib-\Ab$ is invertible,
model \eqref{se} can be written as
\begin{equation}
  \label{see}
  \begin{aligned}
    \yb_i|\xb_i&\sim
    \N_p\left\{\ImA^{-1}\Bb\xb_i,\ImA^{-1}\Sigb\ImA^{-T}\right\}. 
  \end{aligned}
\end{equation}
We define an RGM,
$\GG=(V,E)$, associated with the SEM (\ref{se}) by letting vertices
$V=\{1,\dots,p+q\}$ denote the variables in $(\yb_i,\xb_i)$.
That is, the variables corresponding to $\yb_i$ are indexed
$1,\ldots,p$, and the variables corresponding to $\xb_i$ are indexed
$p+1,\ldots,p+q$.
We allow directed edges terminating in $y$-vertices, $j=1,\ldots,p$,
and undirected edges between $x$-vertices.
Specifically, we draw a directed edge $j'\rightarrow j$ 
from a $y$-vertex $j'=1,\ldots,p$, to another $y$-vertex
$j=1,\ldots,p$ if $a_{jj'} \neq 0$.
And we draw directed edges $h\rightarrow j$ from $x$-vertices
$h=p+1,\ldots,p+q$ to $y$-vertices if $b_{j,h-p} \neq 0$.
Finally, we draw undirected edges $h-h'$ between all
$x$-vertices, $h=p+1,\ldots,p+q$ and $h'=p+1,\ldots,p+q$, $h\neq
h'$.
If needed also the inclusion of undirected edges between $x$-vertices 
could be explicitly modeled. However, we do not pursue
this direction since the focus of this paper is on learning gene
regulations (i.e. directed edges between $y$-vertices). 
  
The distribution of $(\yb_i,\xb_i)$ is global Markov with respect to
the resulting RGM $\GG$ \citep{koster1996markov}, i.e. the presence and absence of edges encode the conditional independence relationships of the vertices. The construction
defines a mapping between SEMs of the type
\eqref{se} and RGMs. In this
paper, $\Bb$ is block diagonal because copy number and methylation
of gene $j$, in principle, only affect the expression of gene $j$. \\
\indent We remark that SEM/RGM also allows for potential causal
  interpretation \citep{koster1996markov,moerkerke2015structural}: the
  presence of directed edges indicates possible direct causal effects
  whereas the absence of directed edges suggests exclusion of such
  possibility. However, the directed edges and cycles should not be
  over-interpreted beyond the scope of SEM without further causal
  assumptions. We refer to
  \cite{robins1999testing,pearl2003causality,goetgeluk2008estimation}
  for detailed discussion on the connection among causal inference,
  graphical models and SEMs.  
\vspace*{-.2in}
\section{hRGMs with known groups}
\label{sec:propmodel}

\subsection{Sampling model}

Suppose the data arises from a heterogeneous population that
is divided into $K$ known groups. 
Let $\yb_k$ and $\xb_k$ be the $n_k\times p$ and $n_k\times
q$ matrices of observations for group $k$, $k=1,\ldots,K$, with
$n=\sum_{k=1}^Kn_k$. We assume that the samples are exchangeable within
each group and the probability model for subject $i$ in group $k$ is
given by 
\begin{equation}
\label{lh}
\yb_{ki} \mid \xb_{ki},\Ak,\Bk,\Sigk \sim
  \N_p\left\{(\Ib-\Ak)^{-1}\Bk\xb_{ki},
         (\Ib-\Ak)^{-1}\Sigk(\Ib-\Ak)^{-T}\right\} 
\end{equation}
for $k=1,\dots,K$ and $i=1,\dots,n_k$ with
$\Ak=\left(a_{jj'}\kk\right)$,
$\Bk=\left(b_{jj'}\kk\right)$ and
$\Sigk=\mbox{diag}\left(\sigk_1,\dots,\sigk_p\right)$. 
Let $\thk = (\Ak,\Bk,\Sigk)$ denote the group-specific parameters. 
By the earlier described mapping between an SEM \eqref{se} and an RGM
model, the $\thk$ define group-specific RGM models $\Gk$. 
The structural zeros of $\Ak$ and $\Bk$ correspond to missing
edges in the RGM $\GG_k$ for group $k$. 

Next we introduce a model feature to induce sparsity 
in $\Ak$ and $\Bk$. We use a non-local prior, defined as follows. First, we expand each entry as 
\begin{equation}
  \label{paex}
  \ak_{jj'}=\atk_{jj'} I(|\atk_{jj'}|>t_{jj'})
  \mbox{~~and~~}
  \bk_{jj'}=\btk_{jj'} I(|\btk_{jj'}|>t_{jj'})
\end{equation}
where $\atk_{jj'}$ and $\btk_{jj'}$ are latent variables
and the threshold parameters $t_{jj'}$ define a minimum
effect sizes of $\ak_{jj'}$ and $\bk_{jj'}$. 
 That is, we represent $\ak_{jj'}$ and $\bk_{jj'}$ by thresholding
latent $\atk_{jj'}$ and $\btk_{jj'}$. A theoretical motivation of
  the construction in (\ref{paex}) is discussed in  \suppl{B}.
The advantages of this thresholding mechanism over standard Bayesian
variable selection framework, for example, a spike-and-slab prior will
become clear when we discuss the priors that link multiple groups.
Notice that $t_{jj'}$
is edge-specific but shared across different groups, which is the key for
inducing the desired dependence of graphs across similar groups and will be discussed in
more detail in Section \ref{sec:plmg}. 

\subsection{Priors linking multiple groups}
\label{sec:plmg}
We first introduce the prior for $\atk_{jj'}$. We assume multivariate
normal priors on
$\tilde{\ab}_{jj'}=\left(\at_{jj'}^{(1)},\dots,\at_{jj'}^{(K)}\right)^T$,  
\begin{eqnarray}
\label{mvnpr}
  \tilde{\ab}_{jj'}\sim p(\tilde{\ab}_{jj'}) = \N(0,\tau_{jj'}\Omegab)
\end{eqnarray}
where $\tau_{jj'}$ is an edge-specific variance component. The
$K\times K$ relatedness matrix $\Omegab=[\Omega_{kk'}]$
links 
edge strength (effect sizes) 
and by the thresholding in \eqref{paex} also edge inclusion
across $K$ different groups, which in
turn also links the graphs across groups. 
The strength of the dependence between groups $k, k'$ is characterized by correlation $\Omega_{kk'}/\sqrt{\Omega_{kk}\Omega_{k'k'}}$. When $\Omega_{kk'}$
is significantly away from zero, $\atk_{jj'}$ and
$\tilde{a}_{jj'}^{(k')}$ are likely to be of similar magnitude
\textit{a priori} and since the thresholding parameters $t_{jj'}$ are
shared across groups, there is a high probability that $\atk_{jj'}$
and $\tilde{a}_{jj'}^{(k')}$ are either both non-zero or both shrunk
to zero. Therefore, graphs $\GG_k$ and $\GG_{k'}$ are more likely to
share common edges. On the other hand, when $\Omega_{kk'}$ is
negligible (i.e. close to 0), groups $k$ and $k'$ are unrelated. We do
not constrain $\Omega_{kk'}$ to be non-negative, which is imposed by
\cite{peterson2015bayesian}. This additional flexibility allows edge
strength to have different signs for different groups, which is
desirable in estimating gene networks because potentially gene
regulations can switch from activation to inactivation across groups
(e.g. case vs control). Note that $\tau_{jj'}$  and $\Omegab$ are
not identifiable because
$\tau_{jj'}\Omegab=c\tau_{jj'}\Omegab/c$ for any $c>0$, which
can be resolved by fixing $\Omega_{11}=1$. The priors for $\btk_{jj'}$
are defined similarly, 
\begin{eqnarray}
\label{eqn:prforb}
\tilde{\bb}_{jj'}=\left(\bt_{jj'}^{(1)},\dots,\bt_{jj'}^{(K)}\right)^T\sim 
     \N(0,\lambda_{jj'}\Omegab).
\end{eqnarray}
The priors in (\ref{mvnpr}) and (\ref{eqn:prforb})
  share the same relatedness matrix $\bm{\Omega}$ that describes
  pairwise similarity of the graphs for each group. The uncertainty of
  each edge strength is captured by the edge-specific variance
  components $\tau_{jj'}$ and $\lambda_{jj'}$.  We assume a simple
  uniform prior for 
	$t_{jj'}\sim\mbox{Unif}(0,b_t)$. We remark that the induced prior for
	$\Ak$ should be restricted to the cone of invertible $\Ib-\Ak$, which
	practically is not a constraint given that any random matrix is almost
	surely invertible \citep{rudelson2008invertibility}.
\vspace*{-.1in}
\subsection{Priors for the relatedness matrix and hyperparameters}
We assign a conjugate inverse-Wishart prior for the relatedness matrix
$\Omegab$ subject to $\Omega_{11}=1$ which controls the relatedness of different groups as
discussed in Section \ref{sec:plmg}, $\Omegab^*\sim \mbox{IW}(\nu,\bm{\Phi})\mbox{~~and~~}\Omegab=\Omegab^*/\Omega^*_{11},$ with degrees of freedom $\nu$ and a scale matrix
$\bm{\Phi}$. Alternatively, one could introduce a hyper
MRF graph $\HH$ and put a G-Wishart prior on
$\Omegab^{-1}\sim \mbox{W}_\HH(\cdot,\cdot)$ if one wants
to learn the conditional independence relationships between
groups. Since in our case study, we only have $K=3$ groups and the
correlations between groups are quite significant (as shown in Section
\ref{sec:bcns}), we are not pursuing this direction in this paper. We
complete the model by assuming conjugate priors $\sigma_j\kk\sim
\mbox{IG}(a_\sigma,b_\sigma)$, $\tau_{jj'}\sim
\mbox{IG}(a_\tau,b_\tau)$ and $\lambda_{jj'}\sim
\mbox{IG}(a_\lambda,b_\lambda)$. For our simulations and case studies,
we use noninformative hyperpriors. But if desired, informative priors can be used, for example, to introduce prior knowledge
about a reference network.
\section{hRGMs with unknown groups}
\label{sec:unkg}
We introduce a clustering approach to split
heterogeneous samples into (unknown) homogeneous groups. 
Under a Bayesian approach, the number $K$ of clusters  need not be
fixed \textit{a priori} and can be inferred from the data. We first
introduce a latent cluster membership indicator $s_i$ 
with $s_i=k$ when subject $i$ belongs to group $k$. Conditional on
$s_i=k$ the likelihood is the same as (\ref{lh}). For  reference
we restate it here, now with conditioning on $s_i$, 
\begin{eqnarray}
\label{lhc}
   \yb_{i} \mid \xb_{i}, \thk,s_i=k \sim 
   \N_p\left\{(\Ib-\Ak)^{-1}\Bk\xb_{i},(\Ib-\Ak)^{-1}\Sigk(\Ib-\Ak)^{-T}\right\}.
\end{eqnarray}
We use the same priors for cluster-specific parameters,
$\tilde{\ab}_{jj'}\sim p(\tilde{\ab}_{jj'}) =
\N(0,\tau_{jj'}\Omegab)$, $\tilde{\bb}_{jj'}\sim  
\N(0,\lambda_{jj'}\Omegab)$and $\sigma_j\kk\sim
\mbox{IG}(a_\sigma,b_\sigma)$
as in Section  \ref{sec:propmodel}. We complete the model with
prior distributions for $s_i$ and $K$. We use a Dirichlet-multinomial (DM) allocation model,
$$
  s_i|\pib,K \sim \mbox{Multinomial}(1,\pi_1,\dots,\pi_K)
  \mbox{ and }
  \pib|K \sim \mbox{Dir}(\eta,\dots,\eta),
$$
and a geometric prior for $K\sim \mbox{Geo}(\rho)$. 
We refer to model (\ref{lhc}) with a DM prior as the hRGM-DM. Alternatively, we consider
nonparametric Bayesian priors that give rise to exchangeable
random partitions such as the Poisson-Dirichlet process, also known as
Pitman-Yor (PY) process priors \citep{pitman1997two}. A PY process
induces a prior distribution on $s_i$'s and $K$ which is characterized
by a (modified) Chinese restaurant process $\mbox{CRP}(\alpha,d)$ with
total mass parameter $\alpha$ and discount parameter $d$, 
\begin{equation}
\label{crp}
  p(s_i=k \mid s_1,\dots,s_{i-1}) \propto \begin{cases}
    {n_k^{(-i)}-d}    & \mbox{for } k=1,\dots,K^{(-i)}\\
    {\alpha+dK^{(-i)}}& \mbox{for } k=K^{(-i)}+1
\end{cases}
\end{equation}
where $n_k^{(-i)}$ and $K^{(-i)}$ are the size of cluster $k$ and the
total number of clusters after removing the $i$th sample. 
The admissible values
for $(\alpha,d)$ are $d\in [0,1)$ with $\alpha>-d$ or $d<0$ with
$\alpha=|md|$ for some integer $m$. The popular Dirichlet process
\citep{ferguson1973bayesian,blackwell1973ferguson} is a special case
of a PY process when $d=0$. The extra parameter $d$ in the PY process makes
it more flexible than the Dirichlet process prior \citep{de2015gibbs}. We
refer to model (\ref{lhc}) with the PY process prior as the hRGM-PY.
\cite{de2015gibbs} and \cite{barrios2013modeling} argue for the PY prior as
more flexible prior for random partitions than DP prior and a parametric DM
prior.

The choice of $\Omegab$ is motivated by the following
consideration. 
The goal is to divide subjects into subpopulations with distinct
networks, rather than induce dependence between clusters.
We therefore set the relatedness matrix
$\Omegab=\mbox{diag}(\Omega_1,\dots,\Omega_K)$ to be diagonal,
i.e. networks are independent across clusters.  

Implementing posterior inference under the hRGM-DM or the hRGM-PY
model is straightforward by MCMC simulation.  Details are provided in
\suppl{C},  including how we report point estimates for networks. 


\section{Case studies}
\label{sec:cs}
We validate inference in
proposed hRGM with extensive simulation studies in \suppl{D}. In this
section, we discuss application of the proposed inference
to three TCGA genomic datasets.
\subsection{p53 pathway across three cancer subtypes}
\label{sec:bcns}
A recent pan-cancer genomic study \citep{hoadley2014multiplatform} has
identified 11 major cancer subtypes based on molecular
characterizations instead of tissue-of-origin. This provides
independent information for clinical prognosis. Although 
the subtypes are correlated
with tissue-of-origin, several distinct cancer types are classified
into common subtypes. For example, head and neck squamous cell
carcinoma (HNSC) and lung squamous cell carcinoma (LUSC) are
classified into subtype ``C2-Squamous-like" by molecular
alterations including p53 whereas invasive breast cancer (BRCA) is
identified as subtype ``C3-BRCA/Luminal" by itself. 

Using inference under the proposed hRGM, we
explore related dependencies of gene networks in the same
three cancer types, HNSC, LUSC and BRCA.
That is, we  construct a gene network for each cancer and 
borrow strength across cancer types adaptively, depending on how
similar the cancers are. We expect the networks of HNSC and LUSC to
have stronger association than those of HNSC and BRCA or of LUSC and
BRCA since HNSC and LUSC belong to the same molecular subtype.

We use the software package TCGA-Assembler \citep{zhu2014tcga} to retrieve mRNA
gene expression, DNA copy number and DNA methylation for HNSC, LUSC
and BRCA  from The Cancer Genome Atlas (TCGA). We focus on genes that
are mapped to the core members of the p53 pathway ($p=10$) which
respond to stresses that 
can cause errors in DNA replication and cell division
\citep{harris2005p53} and plays a critical role in all three cancers
\citep{gasco2002p53,leemans2011molecular,perez2012squamous}. A
more detailed description of these 10 genes appears in \suppl{E}.
We use the same samples of HNSC, LUSC and luminal BRCA that were
analyzed in \cite{hoadley2014multiplatform} and match 
genomic data from different platforms, including methylation
and copy number variation corresponding to each gene. That is, we record $q=2p=20$ DNA
level variables.
The resulting dataset  includes $p+q=30$ variables for 
$n_1=298$ HNSC samples, $n_2=100$ LUSC samples and $n_3=432$ luminal
BRCA samples. We run MCMC simulation for 100,000 iterations, discard
the first 50\% as burn-in and thin the 
chain to every 5th sample. MCMC diagnostics and model checking show no evidence for lack
of practical convergence and lack of fit (see \suppl{E}
for details).

\begin{figure}[h]
	\centering
	\subfigure[HNSC]{\includegraphics[width=0.32\textwidth]{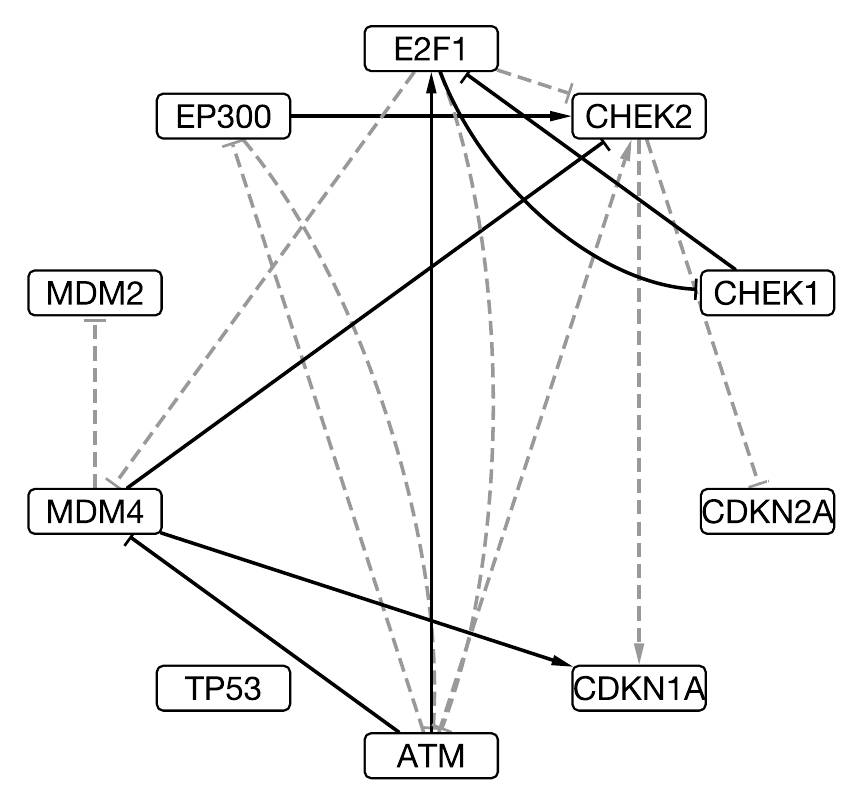}\label{lnet}}
	\subfigure[LUSC]{\includegraphics[width=0.32\textwidth]{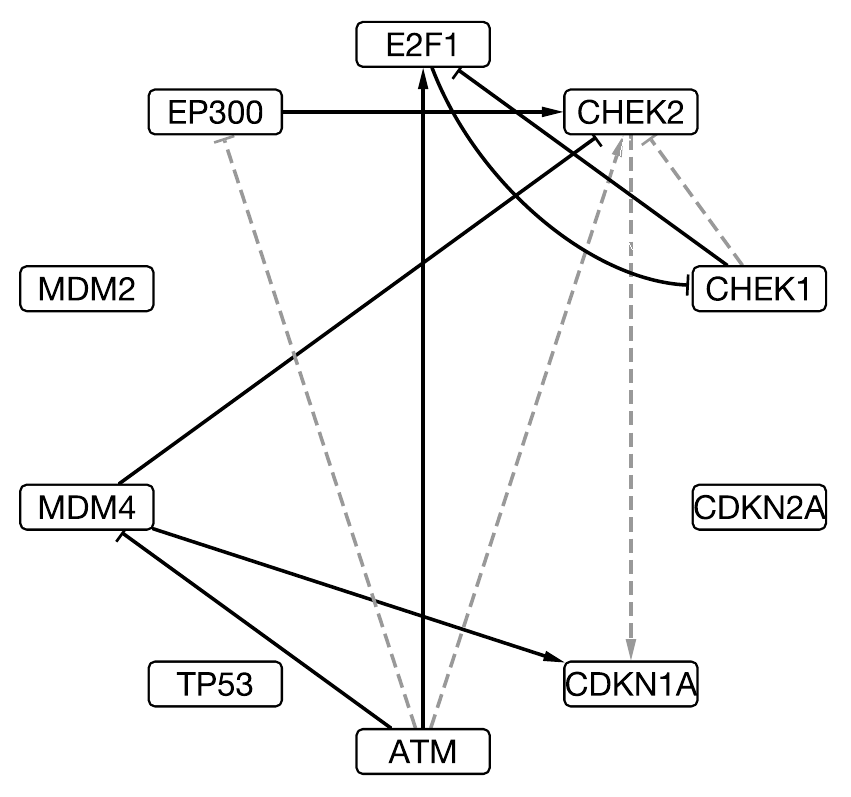}\label{hnet}}
	\subfigure[Luminal BRCA]{\includegraphics[width=0.32\textwidth]{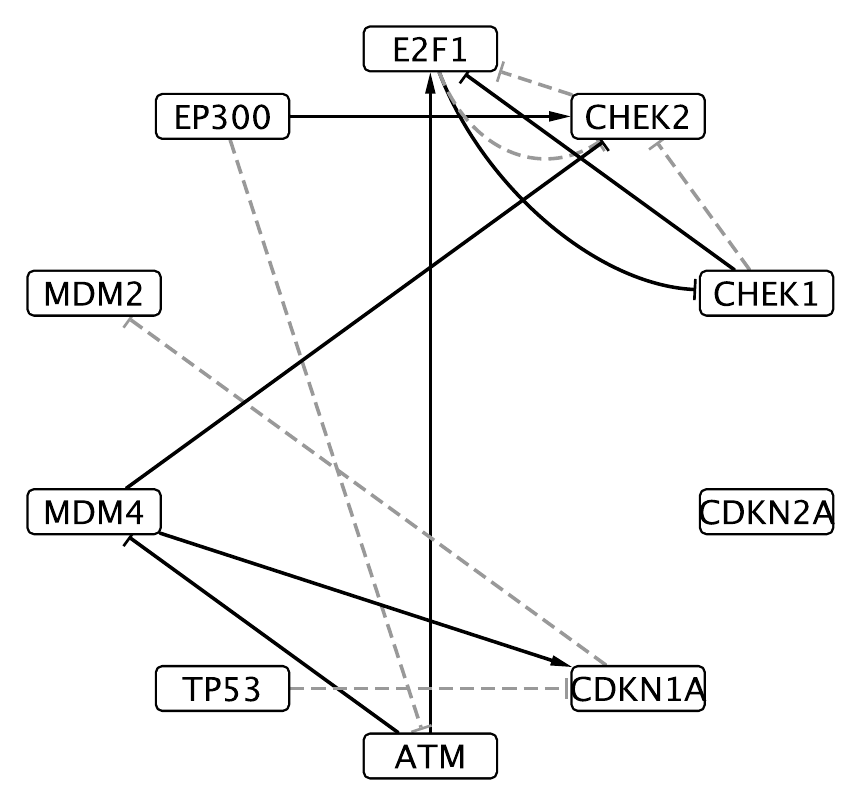}\label{bnet}}
	\caption{Inferred gene networks for HNSC, LUSC and luminal
          BRCA. Shared edges across all subtypes are represented by
          solid lines;  differential edges are shown as dashed
          lines. Arrowheads represent stimulatory interactions,
          whereas horizontal bars denote inhibitory
          regulations. }\label{realnets}	 
\end{figure}

Figure \ref{realnets} shows point estimates for the gene
networks, based on controlling posterior expected 
FDR \citep{newton2004detecting,muller2006fdr} at $1\%$.
The edges between genes and corresponding  copy number
and methylation are omitted for clarity.  These associations are
shown separately in \suppl{E}. 
Most gene expressions are associated
with the corresponding copy number whereas only a few genes are
correlated to their methylation. In Figure \ref{realnets}, 
edges that are shared across subtypes are represented as
solid lines.
Differential edges are represented as dashed lines. Arrowheads denote
stimulatory regulations, whereas small perpendicular
bars indicate inhibitory
regulations. The number of edges in each of three gene networks
and the number of shared edges between each pair of networks are given
in the upper triangular part of Table \ref{table:mat}.
Due to the molecular similarity of HNSC and LUSC, they share many
edges: 10 out of 11 edges in the LUSC network are also found in the HNSC
network. And as expected, BRCA shares fewer, but still a reasonable
number of edges with HNSC and LUSC. 
In summary, inference under the proposed hRGM model 
recognizes difference in association between cancer types and borrows strength
adaptively in a way that confirms the subtypes found by
\cite{hoadley2014multiplatform}.
The estimated levels of adaptive borrowing of strength are reflected in
the estimated relatedness matrix $\widehat{\Omegab}$ (given in the lower triangular part of Table \ref{table:mat}).
The moderate to strong correlations suggest that the joint analysis
under the hRGM is more appropriate than separate inference under
separate RGMs. In fact, separate RGMs detect much fewer edges and
  obtain networks with almost no shared edges. The results are
  provided in \suppl{E}.
\begin{table}[h]
	\caption{Network summaries for HNSC (H), LUSC (L) and BRCA (B). Upper triangular: the number of edges in each of the three gene networks and the number of shared edges between each pair of networks. Lower triangular: the estimated relatedness matrix $\widehat{\Omegab}$.}
	\begin{center}
		\begin{tabular}{cccc}
			\hline\hline
			&H&L&B\\
			H&16&10&9\\
			L&0.77&11&8\\
			B&0.63&0.57&13\\\hline
		\end{tabular}
	\end{center}
	\label{table:mat}
\end{table}
A noticeable feature in all three estimated networks in Figure
\ref{realnets} is the central role of E2F1. 
E2F1 is an important transcription factor gene across cancer types which
interacts with multiple genes in all networks. Such highly connected
genes are known as hub genes and are often more involved in
multiple regulatory activities than non-hub genes. In fact, E2F1 has a
pivotal role in controlling cell cycle progression and induces
apoptosis. It has been found that E2F1 is mutated in most, if not all,
human tumors \citep{polager2009p53}. Our finding that E2F1 plays
important roles in all networks is consistent with the fact that hub
genes are more likely to be conserved across species, diseases and
tumor (sub-)types \citep{casci2006network}. In addition, some edges
that we find in our analysis are well studied in the biological
literature. For example, ATM phosphorylates and stabilizes E2F1 in
response to various stresses including DNA damage. Our study confirms
this positive regulatory relationship across all cancers. 

\subsection{KEGG colorectal cancer pathway cross two cancers}
\begin{figure}[h]
	\centering
	\subfigure[COAD full network]{\includegraphics[width=0.49\textwidth]{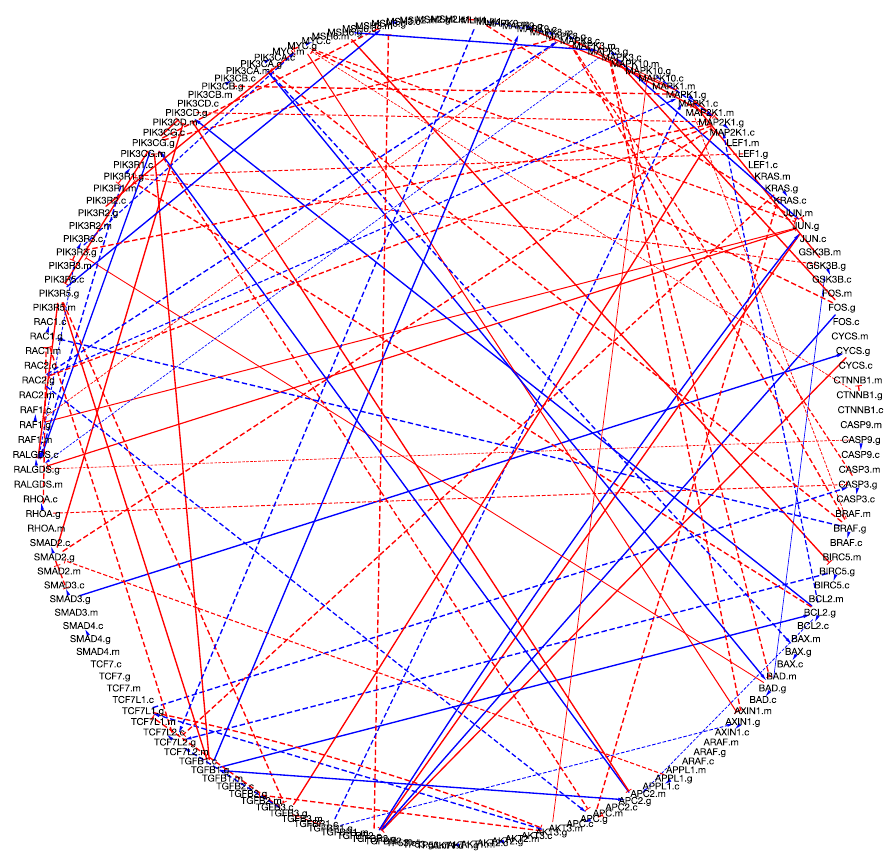}\label{cf}}
	\subfigure[READ full network]{\includegraphics[width=0.49\textwidth]{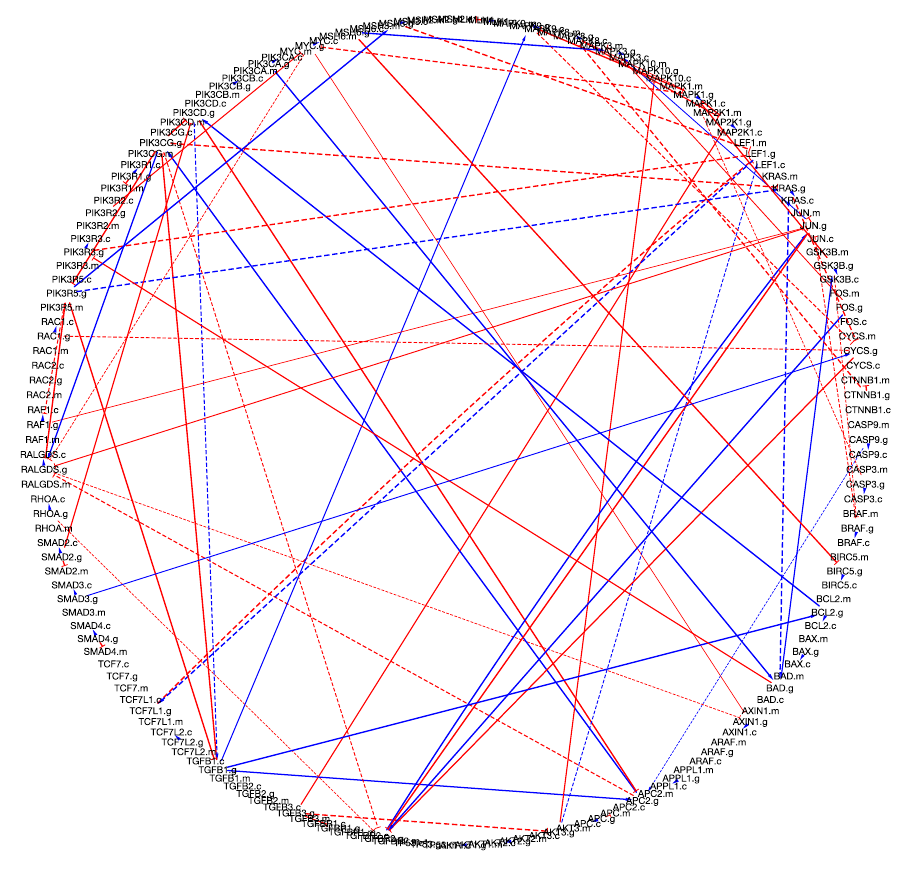}\label{rf}}		\subfigure[COAD gene network]{\includegraphics[width=0.49\textwidth]{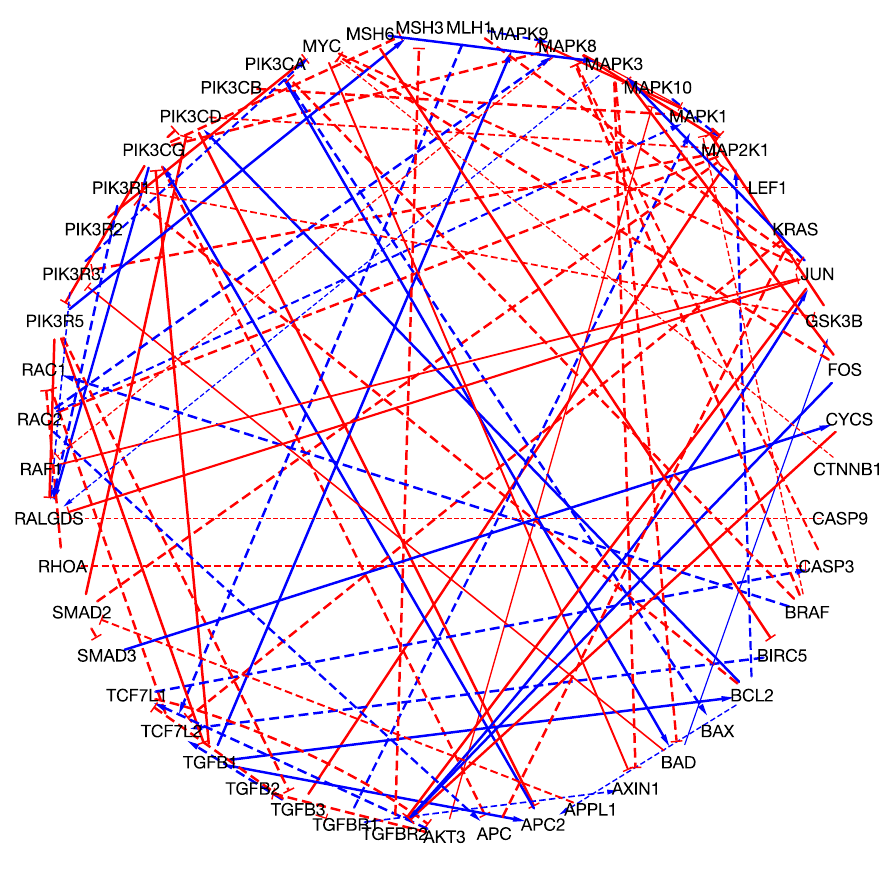}\label{cg}}
	\subfigure[READ gene network]{\includegraphics[width=0.49\textwidth]{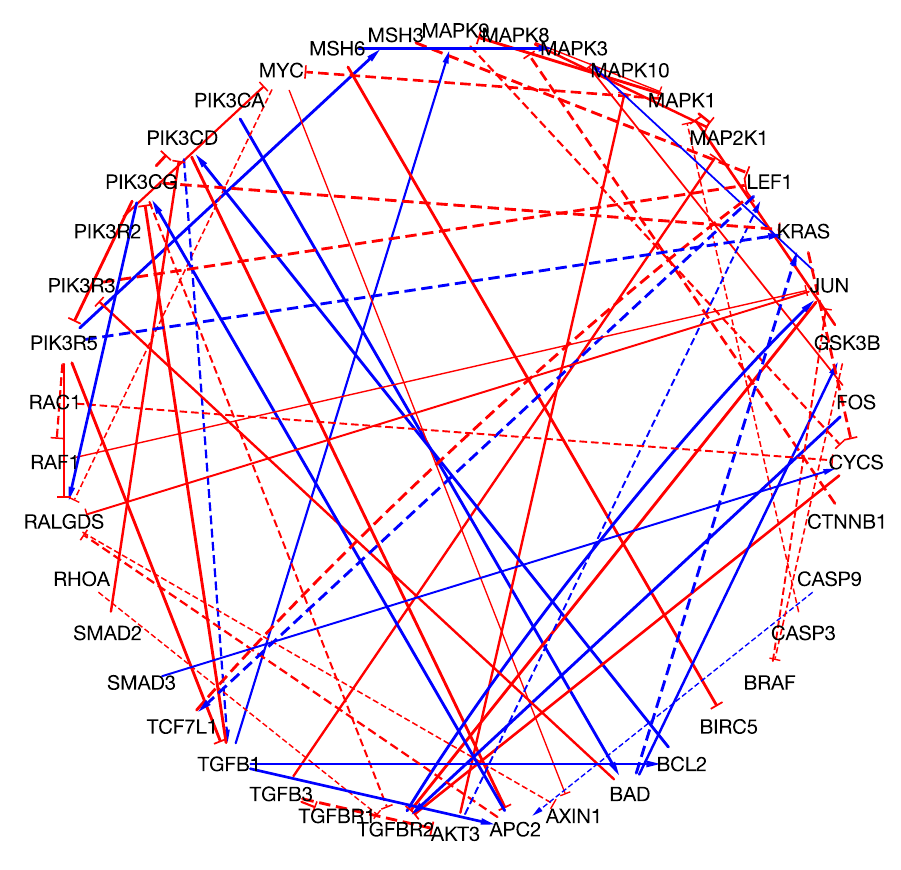}\label{rg}}
	\caption{Inferred networks for COAD and READ. Shared edges across cancer types are represented by
		solid lines;  differential edges are shown as dashed
		lines. Edges with arrowheads represent stimulatory interactions,
		whereas edges with horizontal bars denote inhibitory
		regulations. (a) Full network in COAD with copy number, methylation and gene expression. (b) Full network in READ. (c) Gene network in COAD. (d) Gene network in READ. In (a) and (b),  the suffixes represent: c=copy number, m=methylation and g=gene. In (c) and (d), disconnected genes are omitted.}\label{Colo}
\end{figure}

In this application, we investigate the genomic patterns of colon
adenocarcinoma (COAD) and rectum adenocarcinoma (READ). COAD and READ
are closely related cancers in terms of the copy number alteration,
DNA methylation and gene expression patterns
\citep{cancer2012comprehensive}. However, they are also known
to exhibit substantial
difference in genetic characterization \citep{frattini2004different},
the process of carcinogenesis \citep{kapiteijn2001mechanisms} and
chemosensitivity \citep{kornmann2014differences}. We download DNA copy
number, methylation and mRNA gene expression using TCGA-Assembler
\citep{zhu2014tcga} for genes that are mapped to the KEGG colorectal
cancer pathway (\url{http://www.genome.jp/}). The dataset contains
$n_1=276$ COAD samples, $n_2=91$ READ samples and totally 174 genomic
measurements (58 in each of copy number, methylation and gene
expression).\\ 
\indent The full networks with 174 vertices for COAD and READ are
shown in Figures \ref{cf} and \ref{rf}, respectively. They share 77
edges which are represented by solid lines. Each cancer has moderate
number of differential edges as well, indicated by dashed lines: COAD
has 60 distinctive edges whereas READ has 27 unique edges. \\ 
\indent For better visualization, we focus on the gene networks
(i.e. without copy number and methylation) in Figures \ref{cg} and
\ref{rg}. We find in COAD, PIK3R1 inactivates GSK3B. The biological
literature confirms this finding that PIK3R1 phosphorylates and
activates AKT which in turn phosphorylates and inactivates GSK3B
\citep{danielsen2015portrait,lin2015pik3r1}. Surprisingly, however, this link is missing in the READ gene network. This unexpected result deserves further experimental validation.\\
\indent PIK3CG, a catalytic subunit of phosphatidylinositide 3-OH
kinase, seems to play a critical role in both COAD and READ: it has 8
and 7 links to other genes, respectively. In fact, down-regulation of
PIK3CG expression inhibits PI3K-AKT signaling pathway responsible for
carcinogenesis and progression of colorectal cancers, which occurs
frequently in primary colorectal cancers \citep{semba2002down}. 

\subsection{Clustering breast cancer subtypes}
\label{sec:bcug}
Breast cancer is known to be a highly
heterogeneous disease. Based on three biomarkers (ER/PR/HER2), breast cancer is  traditionally  classified into
three groups: luminal, HER2 and basal.  However, this
classification system is also known  to be suboptimal and 
 in need of improvement 
for better diagnostics and prognostics 
\citep{di2015new}. 

Instead of three biomarkers, we consider $p=10$ core genes of a critical pathway, RAS-MAPK, which transmits and amplifies signals involved in cell
proliferation and cell death in breast cancer
\citep{santen2002role}. These 10 genes  form the  key
  component of the RAS-MAPK pathway  known as  the MAPK
  cascade which is explained in detail in \suppl{E}.
We use inference
under the proposed hRGM-DM to find subtypes based on  
breast cancer data ($n=720$ samples) retrieved
from TCGA as described in Section \ref{sec:bcns}. Our goal is to
simultaneously cluster patients and estimate gene networks for each
cluster. We  implement inference using MCMC posterior simulation,
using the same  setup as in Section
\ref{sec:bcns}. MCMC diagnostics and model checking show no
  evidence for lack of practical convergence and lack of fit (see
  \suppl{E} for details).

\begin{figure}[h]
	\centering
\subfigure[hRGM-DM]{\includegraphics[width=.24\textwidth]{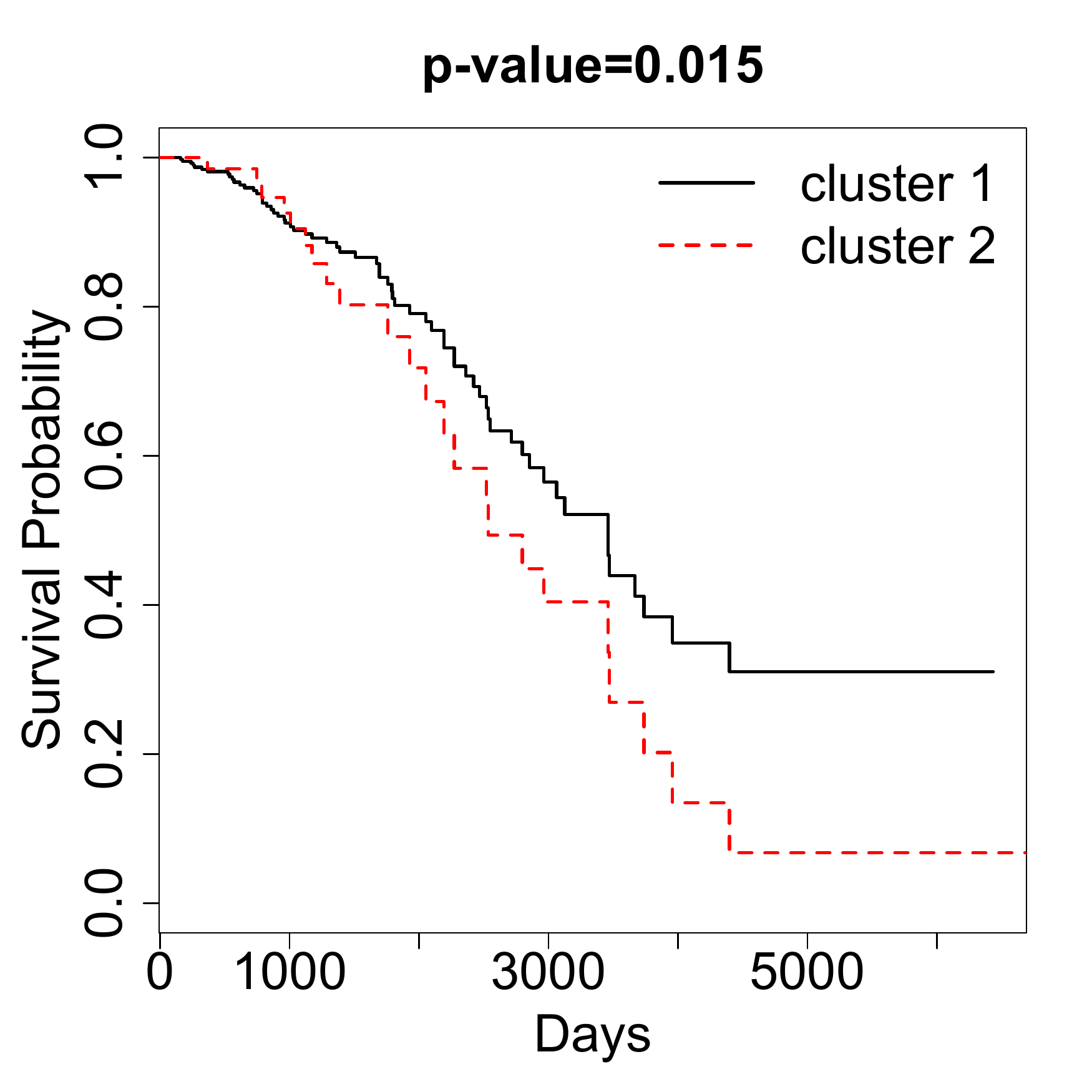}\label{fig:km}}	
\subfigure[RLD]{\includegraphics[width=.24\textwidth]{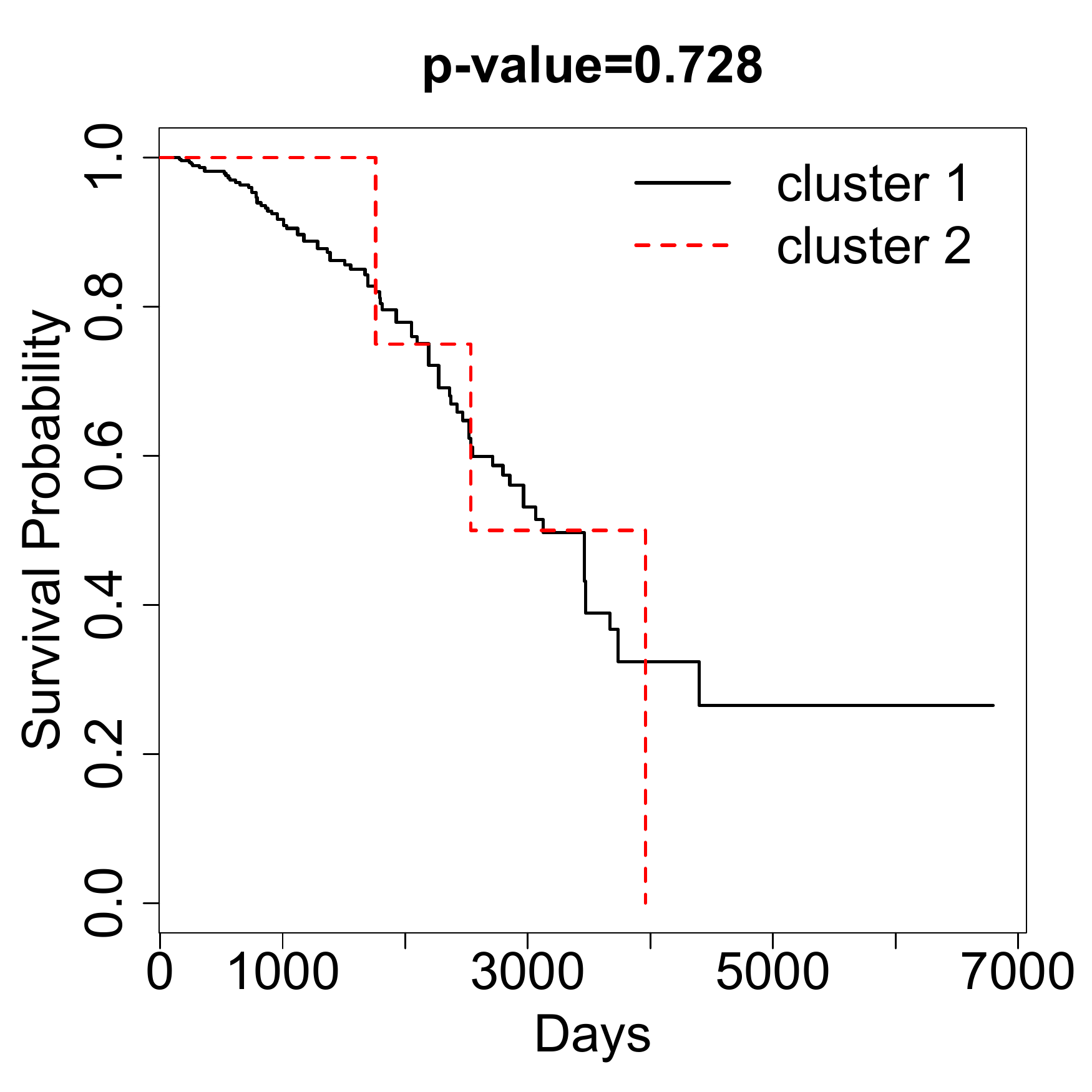}\label{fig:kmrld}}
\subfigure[K-means]{\includegraphics[width=.24\textwidth]{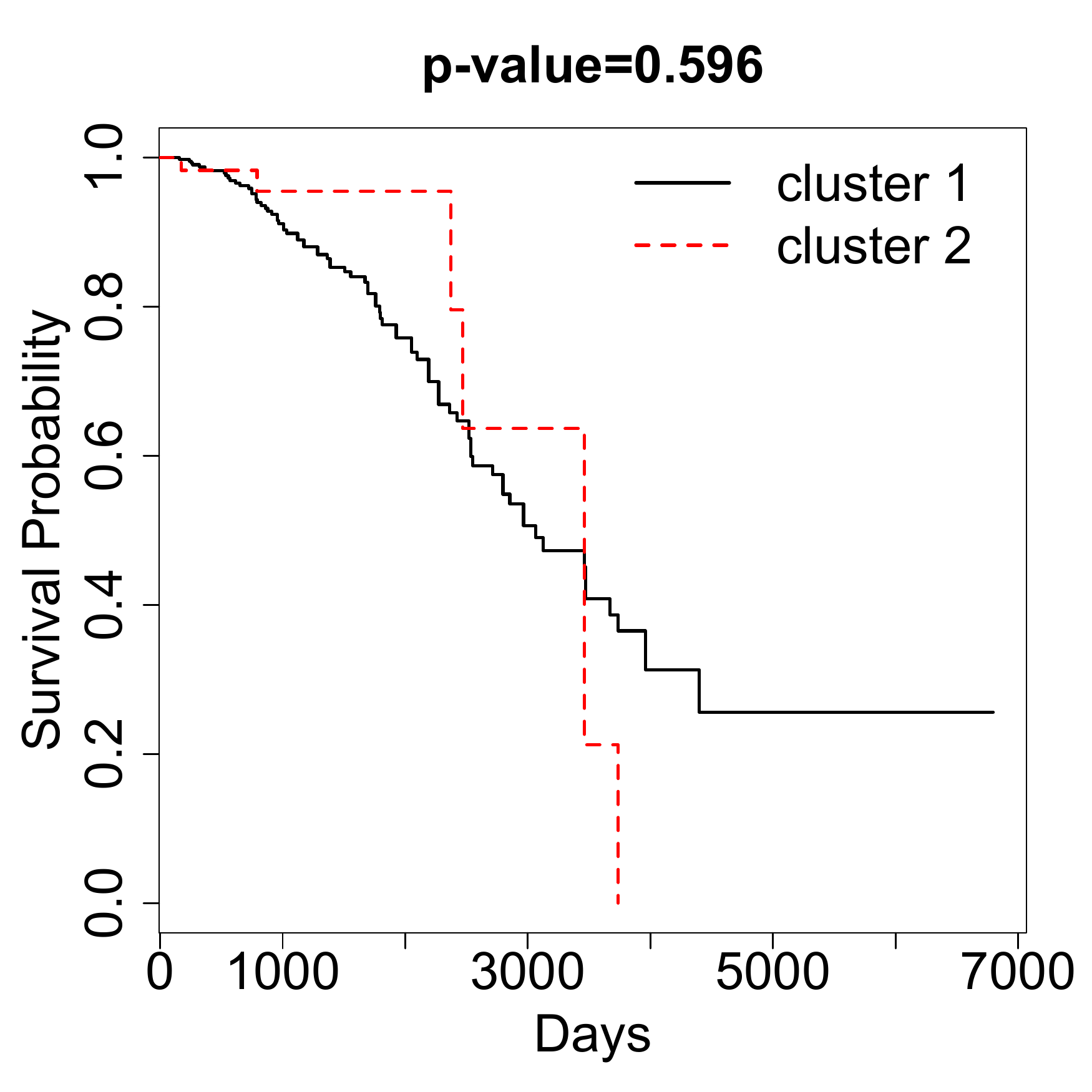}\label{fig:km2}}
\subfigure[Subtypes]{\includegraphics[width=.24\textwidth]{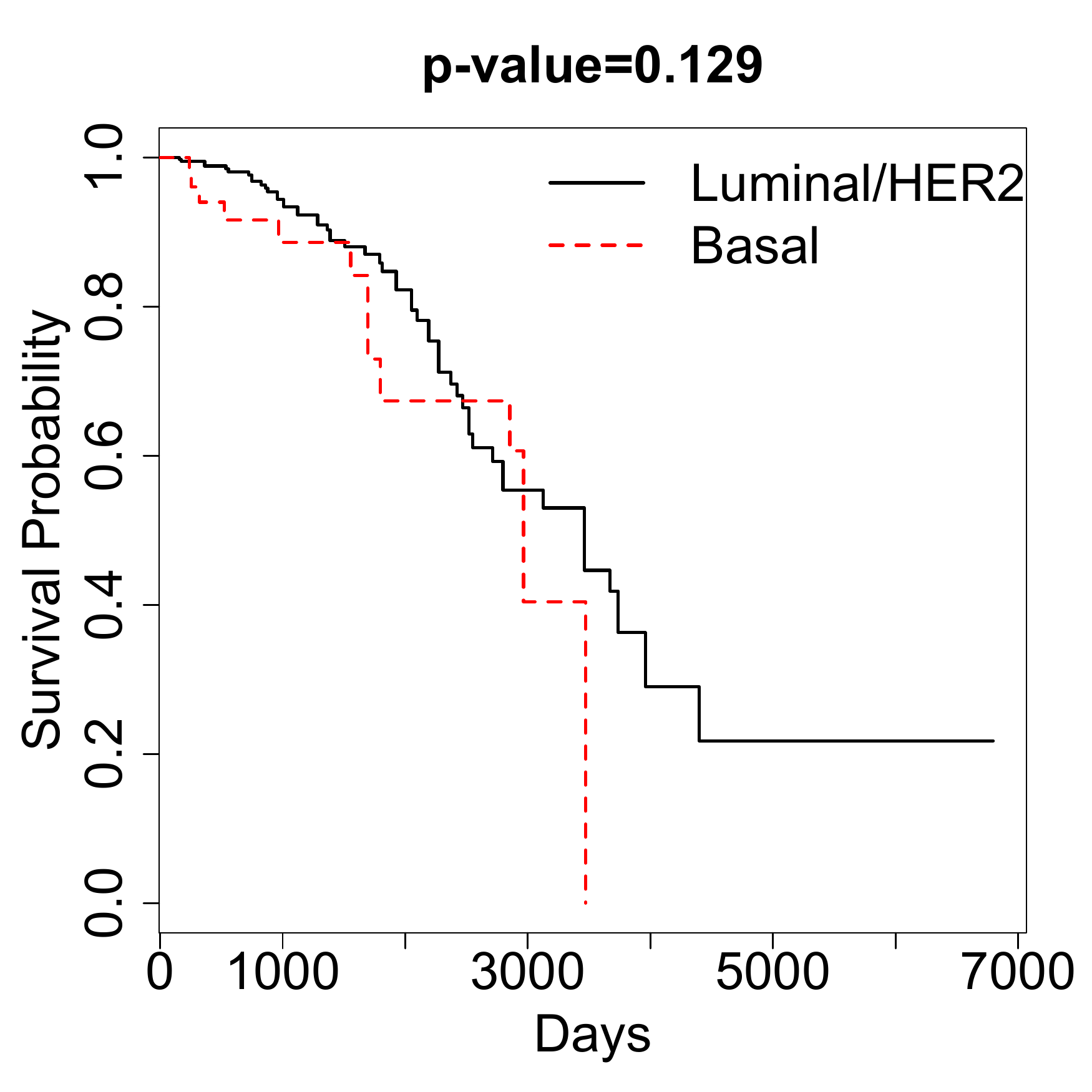}\label{fig:kmsub}}
	\caption{Kaplan-Meier estimators for cluster 1 (solid) and cluster 2 (dashed). P-values of log-rank test are shown on the top of each figure.}\label{fig:kms}
\end{figure}

We identify two major clusters with 600 and 107 samples. 
 To explore the clinical relevance of the reported clusters we
evaluate Kaplan-Meier (KM) estimates of patients' survivals for the two
clusters. These  are shown in Figure \ref{fig:km}. We find a substantial
difference in median survival between the two clusters: a difference
of 928 days. For comparison, 
 we also apply RLD \citep{rodriguez2011sparse}
and the K-means algorithm with $K=2$ to the same data including gene
expression, copy number and methylation. Their KM estimators are 
shown in Figures \ref{fig:kmrld} and \ref{fig:km2}. RLD finds one
major cluster with 687 patients. The second largest cluster has only
22 patients. The difference of median survival is 120.5 days between
the two clusters. The two clusters identified by K-means have 635 and
85 samples with median difference in survival of 399
days. In addition, we 
also compute KM estimates, shown in Figure \ref{fig:kmsub}, for the
luminal/HER2 group versus basal group;  the latter usually has much
worse prognosis. The difference of median survival is 497 days. 

\begin{figure}[h]
	\centering
	\subfigure[Cluster 1]{\includegraphics[width=0.4\textwidth]{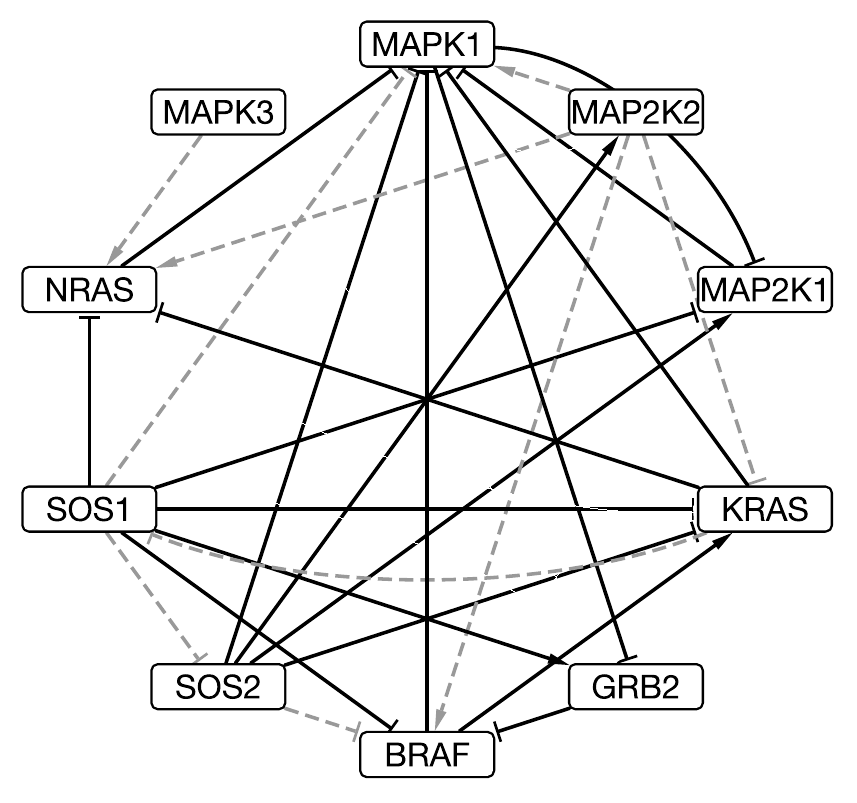}\label{net1}}
	\subfigure[Cluster 2]{\includegraphics[width=0.4\textwidth]{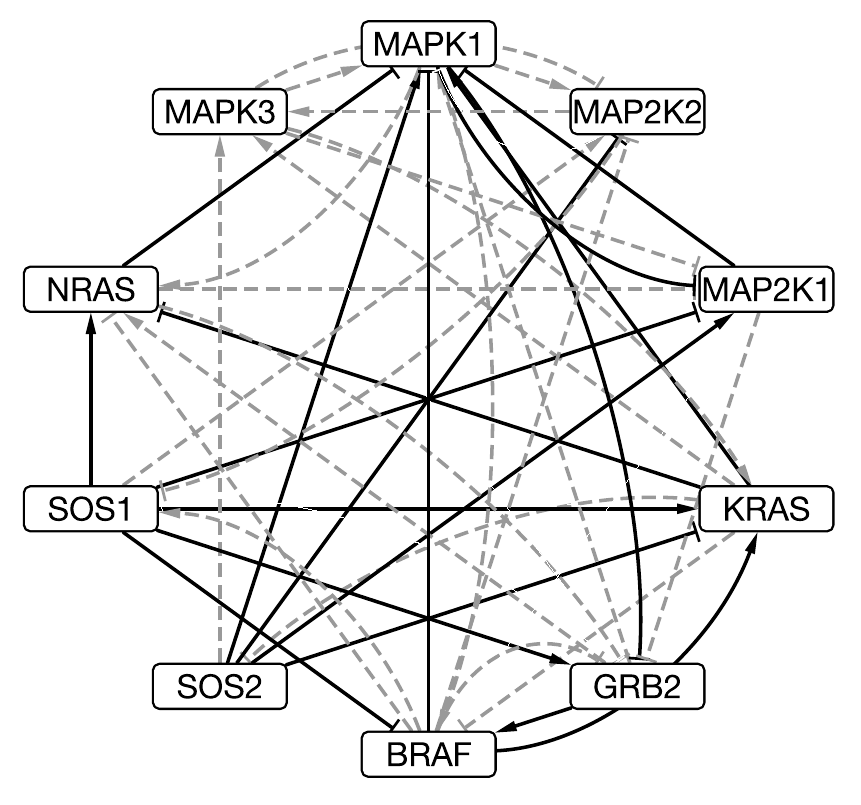}\label{net2}}	
	\subfigure[Feedback loop for cluster 2]{\includegraphics[width=.75\textwidth]{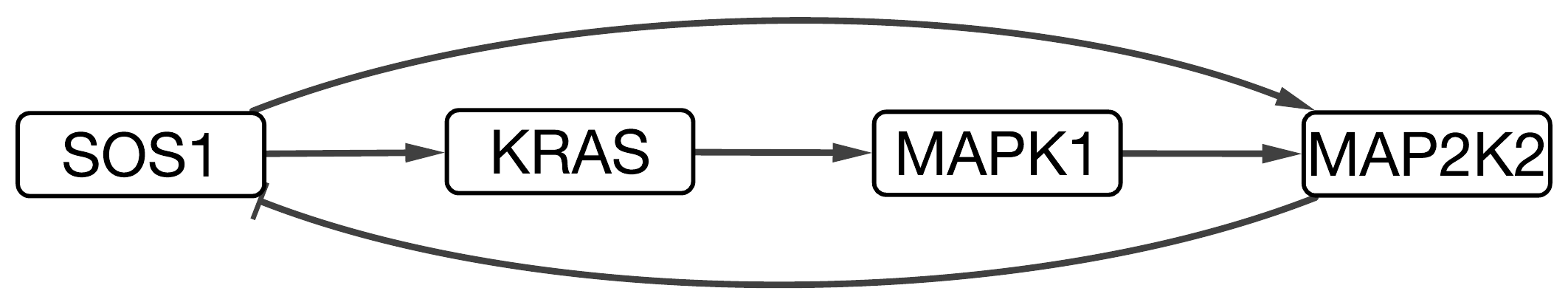}\label{fig:casc}}
	\caption{Inferred gene networks for breast cancer. Top two figures: networks for clusters 1 and 2 with shared edges represented by solid lines and differential edges represented by dashed lines. Bottom figure: negative feedback loop of gene network for cluster 2.  Arrowheads represent stimulatory interactions, whereas horizontal bars denote inhibitory regulations.}\label{clnets}	
\end{figure}

To formally compare the clinical relevance in terms of
significantly different overall survival for the reported clusters,  we
carried out log-rank tests (p-values shown in Figure \ref{fig:kms}) for the difference of the corresponding
survival distributions.
In fact, only hRGM-DM identifies clusters that
have significantly different survival distributions. 
The hRGM-DM model detects the most distinctive clusters in terms of
prognosis.
Such clusters or subtypes  
may be useful in refining subtypes of breast cancer and developing new
treatment options. In addition, we quantify the difference between
  clusters estimated from hRGM-DM, RLD and K-means by the adjusted
  Rand index and the adjusted variation of information
  \citep{meilua2003comparing}  in the \suppl{E}.
  Similarly to what the KM estimates suggest, the clusters obtained
  from each method are quite different.  \\  
\indent In terms of the estimated network structure, we find that all
genes are associated with their respective copy number for both
clusters except for SOS1 and KRAS in cluster 1 and SOS1 and MAPK3 in
cluster 2. In addition, NRAS and KRAS in both clusters and SOS2 and
BRAF in cluster 2 are found to be associated with their
methylation. 
Figure \ref{clnets} shows the estimated gene networks for the two
clusters 
when controlling posterior expected FDR at $1\%$.
As before, shared edges across both clusters  are represented by
solid lines and differential edges are shown as dashed
lines. Arrowheads denote stimulatory regulations, whereas horizontal
bars indicate inhibitory regulations. We find 27 edges for cluster 1
and 41 edges for cluster 2. The two networks share 18 edges but are
otherwise quite different from each other. For example, the well-known
cascade SOS1$\rightarrow$KRAS$\rightarrow$MAPK1$\rightarrow$MAP2K2
\citep{santen2002role} is found in cluster 2 but not in cluster
1. Furthermore, we find MAP2K2 inhibits SOS1 in cluster 2, which
completes the negative feedback loop, a commonly observed motif in
gene network \citep{krishna2006structure}, as shown in Figure
\ref{fig:casc}. Interestingly, it is recently discovered that MAP2K2
phosphorylates and inhibits SOS, therefore reducing MAP2K2 activation
\citep{mendoza2011ras}. 

\section{Discussion}
\label{sec:disc}
We have developed the hRGM(-DM,-PY) model for inference on gene networks
for heterogeneous samples. The hRGM 
connects graphs across known groups by introducing correlation on edge
strength and edge inclusion. 
 In the case of unknown groups, the hRGM together with
a Dirichlet-multinomial or Pitman-Yor process prior for the cluster
arrangement allows to learn the unknown groups 
and estimate group-specific networks at the same time. Bayesian model
selection  is implemented with a thresholding prior and used to
obtain sparse networks. Simulation studies demonstrate the advantages
of hRGM over a  comparable approach with separate inference for
group-specific networks. 
In the first application, we find both common and differential network
structures for different cancer types. In the second application, we
are able to identify clusters that differ significantly in network
structures.  The clinical significance of the discovered clusters
is validated by significantly different cluster-specific
survival functions. Inference provides important information for refining subtypes of breast cancer and for developing new therapeutic
options.\\
\indent In this paper, we only consider continuous random
  variables and assume normal distributions. This is
  because the mapping between an RGM and an SEM is only valid for
  Gaussian distributions \citep{koster1996markov}. Although
  we have empirically demonstrated the robustness of our method to the
  distribution misspecification, it would be desirable to extend the model to non-Gaussian data.
  For example, one may want to integrate data on single nucleotide
  polymorphisms (SNPs) and mutation status (MUs) into the gene
  network reconstruction. However, SNP and MU are recorded as discrete
  variables.  
  One possible implementation is the use of latent Gaussian variables.
  In principle, RGMs can be embedded in a more general exponential
  family graphical model framework \citep{yang2015graphical}. However,
  without the mapping between RGM and SEM, computation is a major
  bottleneck for such extensions. 
  
 \vspace*{-.1in}
\section{Supplementary Materials}

Web Appendices, Tables, and Figures referenced in Sections \ref{sec:back}, \ref{sec:unkg} and \ref{sec:cs} and Matlab program implementing our methods, are available with this paper at the Biometrics website on Wiley Online Library.

 \vspace*{-.2in}
 \bibliography{mrgm_ref}

\label{lastpage}

\end{document}